\documentclass[aps,prb,twocolumn,showpacs]{revtex4-1}
\usepackage{graphicx}
\usepackage{rotating}
\usepackage{amsfonts,amsmath,color}
\usepackage{amssymb}
\usepackage{psfrag}

\newcommand{\To}{{\mbox{$T_0$}}}

\newcommand{\vo}{{\mbox{$v_0$}}}
\newcommand{\vn}{{\mbox{$v_n$}}}
\newcommand{\vs}{{\mbox{$v_s$}}}
\newcommand{\Vs}{{\mbox{$V_s$}}}
\newcommand{\vns}{{\mbox{$v_{ns}$}}}
\newcommand{\rhon}{{\mbox{$\rho_n$}}}
\newcommand{\rhos}{{\mbox{$\rho_s$}}}
\newcommand{\vvs}{{\bf {v}}_s}
\newcommand{\vvn}{{\bf {v}}_n}

\newcommand{\VVs}{{\bf {V}}_s}
\newcommand{\sss}{\bf {s}}
\newcommand{\rr}{\bf {r}}
\newcommand{\dr}{\rm {d}}

\newcommand{\bnabla}{{\mbox{\boldmath $\nabla$}}}

\newcommand{\oder}[2]{\frac{{\rm d} #1}{{\rm d} #2}}

\begin{document}

\title{Superfluid turbulence driven by cylindrically symmetric thermal counterflow}

\author{E.~Rickinson}
\affiliation{Joint Quantum Centre Durham-Newcastle, and School of Mathematics, Statistics and Physics, Newcastle University, Newcastle upon Tyne, NE1 7RU, UK}

\author{C.\,F.~Barenghi}
\affiliation{Joint Quantum Centre Durham-Newcastle, and School of Mathematics, Statistics and Physics, Newcastle University, Newcastle upon Tyne, NE1 7RU, UK}

\author{Y.\,A.~Sergeev}
\affiliation{Joint Quantum Centre Durham-Newcastle, and School of Mathematics, Statistics and Physics, Newcastle University, Newcastle upon Tyne, NE1 7RU, UK}

\author{A.\,W.~Baggaley}
\affiliation{Joint Quantum Centre Durham-Newcastle, and School of Mathematics, Statistics and Physics, Newcastle University, Newcastle upon Tyne, NE1 7RU, UK}

\date {\today}

\begin{abstract}
We show by direct numerical simulations
that the turbulence generated by steadily heating a long cylinder
immersed in helium~II is strongly inhomogeneous and consists of a 
dense turbulent layer of quantized vortices localized around the cylinder.
We analyse the properties of this superfluid turbulence in terms of
radial distribution of the vortex line density and the anisotropy and we
compare these properties to the better known properties 
of homogeneous counterflow turbulence in channels.
\end{abstract}

\maketitle

\section{Introduction}
\label{Intro}

Since the early experiments of Vinen \cite{Vinen},
superfluid turbulence has been typically studied in
a long channel which is closed at one end and connected
to the helium bath at the other end.
At the closed end, an electrical resistor steadily
dissipates a known heat flux which
drives helium's two components in opposite directions:
the normal fluid towards the bath and the superfluid towards the
resistor; this motion is called thermal counterflow.
 If the applied heat flux exceeds a critical
value, an extra thermal resistance is observed, caused by 
the appearance of a turbulent tangle of quantised vortex lines which
limit the heat-conducting properties of liquid helium. 
Following Vinen's work, thermal
counterflow has been the subject of many experiments and numerical
simulations which have revealed the nature and the dynamics
of turbulent vortex lines.
Recently-investigated aspects of the problem 
include the Lagrangian velocity statistics of tracer particles 
which are used to
visualize the turbulence
\cite{Svancara2018,Mastracci2019},
the coupled dynamics of normal fluid and vortex lines \cite{Yui2018}, 
the comparison with ordinary turbulence \cite{Biferale2019}, and the
effects induced by the channel's walls \cite{BaggaleyLaurie2015}.
In the last case the density of vortex lines is not spatially uniform;
indeed inhomogeneous superfluid turbulence is still poorly understood.
 
This work is concerned with perhaps the simplest configuration
of inhomogeneous superfluid turbulence:
steady radial counterflow around a long heated cylinder.
This flow is simple to set-up in the laboratory, requiring only
a thin metal wire across a cell containing liquid helium
through which an electrical current
dissipates a known heat flux into the surrounding liquid. 
The natural question which we address 
is whether the superfluid turbulence around the heated cylinder differs 
from the standard case studied by Vinen. In particular we want to find
whether, and under what conditions, the local vortex line density 
achieves any statistically steady state, and 
determine its radial distribution. Besides turbulence, the
problem of the heated cylinder has motivations of engineering heat transfer and 
applications such as hot-wire anemometry \cite{Duri2015} in liquid helium
\cite{note0}.

\section{Formulation of the problem and numerical method}
\label{sec:formulation}

We numerically model the thermal counterflow generated by a heated, 
infinitely long, cylinder of radius $a$ immersed in liquid helium~II. 
We assume that the cylinder generates a constant heat flux, $q$, 
which determines  the radial normal velocity $\vo=\vn(a)$ 
at the cylinder's surface $r=a$ (hereafter $r$ is the radial coordinate).

We assume that the temperature of helium is uniform throughout the whole 
flow domain, $r>a$, so that the normal and superfluid densities and 
all other thermodynamic quantities are constant. First
we consider the normal ($\vvn$) and superfluid ($\VVs$) velocity 
distributions in the case where there is no superfluid turbulence.
In the thermal counterflow generated by the heated surface of the
cylinder, the normal fluid moves radially out with positive radial velocity 
$\vn=a\vo/r$ taking heat away (hereafter the subscript $r$ in the radial 
components of $\vvn$ and $\VVs$ is omitted). In the steady-state flow regime, 
the counterflow condition

\begin{equation}
\rhon\vvn+\rhos\VVs={\bf 0}\,,
\label{eq:counterflow}
\end{equation}

\noindent
(where $\rhon$ and $\rhos$ are respectively
the normal and superfluid densities) yields the following
radial superfluid velocity:

\begin{equation}
\Vs=-\rhon a\vo/(\rhos r),
\label{eq:Vs}
\end{equation}

\noindent
where the minus sign means that $\VVs$ points radially inwards.

Hereafter we consider the case in which helium~II becomes
turbulent. For the sake of simplicity, we assume that the 
driving velocity $\vo$ is large enough 
that superfluid vortex lines are generated, but not so large that
the normal fluid becomes turbulent. In other words,
we assume the so-called 
T1 regime~\cite{Tough} of counterflow turbulence. We treat the 
superfluid velocity, $\VVs$, which enforces the counterflow 
condition~(\ref{eq:counterflow}) and therefore is radially distributed 
according to Eq.~(\ref{eq:Vs}), as the externally applied superflow;
in this way, in the presence of the turbulent vortex tangle, 
the total superfluid velocity $\vvs$ can be decomposed 
as $\vvs=\vvs^i+\VVs$, where $\vvs^i$ is the self-induced velocity 
generated by the vortex tangle. In the framework of the vortex 
filament method, we model quantum vortex filaments 
as infinitesimally thin space curves $\sss(\xi,\,t)$ 
which move according to the Schwarz equation \cite{Schwarz1988}

\begin{equation}
\oder{\sss}{t}=\vvs+\alpha\sss'\times(\vvn-\vvs)-\alpha'\sss'\times[\sss'\times(\vvn-\vvs)],
\label{eq:Schwarz}
\end{equation}

\noindent
where $t$ is time, $\alpha$ and $\alpha'$ are 
dimensionless temperature-dependent 
friction coefficients~\cite{Donnelly-Barenghi}, 
$\sss'=\dr\sss/\dr\xi$ is the unit tangent vector at the point $\sss$, 
and $\xi$ is the arc length.

At the point $\sss$, the self-induced velocity is given by the Biot-Savart 
law  \cite{Adachi,BagBar2012} 

\begin{equation}
\vvs^i=-\frac{\kappa}{4\pi}\oint_{\cal L}\frac{(\sss-\rr)}{\vert\sss-\rr\vert^3}\times\dr\rr,
\label{eq:BS}
\end{equation}

\noindent
where $\kappa=9.97\times10^{-4}\,{\rm cm^2/s}$ is the quantum of circulation, 
and the line integral extends over the entire vortex configuration $\cal L$.

We numerically simulate the emergence and evolution of the vortex tangle 
for a cylinder of given radius $a$ and
normal fluid velocity $\vo$ at the cylinder's surface.
In all simulations reported here
we assume the values $a=0.1~\rm cm$ and
$\vo=0.6~\rm cm/s$. We also assume that, in the bulk, the temperature
of the liquid helium is $T=1.3~\rm K$.
At this temperature, the normal fluid
and superfluid densities are $\rhon=6.522\times10^{-3}\,{\rm g/cm^3}$ 
and $\rhos=0.1386\,{\rm g/cm^3}$ respectively, and the mutual friction 
coefficients are $\alpha=0.034$ and $\alpha'=1.383\times10^{-2}$.

Our calculation is performed in a 
domain open in the direction orthogonal to the cylinder and periodic 
in the coordinate $z$ along the axis of cylinder, 
with a period of $0.2~\rm cm$. 
The vortex lines are discretized 
by Lagrangian points $\sss_j$ for $j=1,\dots,N$ held at minimum 
separation $\Delta\xi=2\times10^{-3}$~cm; 
the Schwarz equation~(\ref{eq:Schwarz}) is time-stepped
using a fourth-order Adams-Bashforth method with  typical 
time step $\Delta t=5\times10^{-5}$~s.
The de-singularization of the Biot-Savart integrals and
the technique to numerically perform vortex reconnections
when vortex lines approach each other sufficiently close are all 
described in Refs.~\cite{BagBar2012,Baggaley2012,Laurie-Baggaley2015}. 
The number of Lagrangian points changes with time and becomes
very large (typically of the order of $N \approx 10^5$)
in the final statistically steady-state regime of turbulence.
To reduce the computation time, 
during the initial transient from time $t=0$ to the time $t_s$
when the tangle is still dilute and the vortex length builds up,
the Biot-Savart integral is approximated by the Local Induction 
Approximation (LIA)~\cite{Schwarz1978}. 
At the somewhat arbitrary moment $t_s$, when the tangle becomes dense 
in the region adjacent to the cylinder's surface \cite{note1},
we switch from the LIA to the full Biot-Savart integral in the tree-algorithm 
approximation (the switching criterion will be described
in section IV). The tree-algorithm 
(described and tested in Ref.~\cite{BagBar2012}), 
is clearly suitable for random tangles such     
as ours, where contributions to the velocity field at a point
arising from far-away vortex lines are less important than contributions
arising from vortex lines in the immediate neighbourhood of that point.

The initial transient under LIA does 
not affect our final results for the following reason.  
The characteristic time scale for the Biot-Savart interaction between
the vortex lines is of the order of the eddy turnover time
$\tau \approx \ell/u_{\ell} \approx \ell^2/\kappa$ where $\ell$ is the
distance between vortices in a given region. Taking typical values
$L \approx 1000$ to $6000~\rm cm^{-2}$ for the vortex line density 
in the saturated regime at distances $a<r<5a$, %
we find $\tau \approx 0.2$ to $1~\rm s$,
which is much shorter than the time-scale $\approx 23~\rm s$
of the evolution computed under the Biot-Savart law.

The boundary conditions at the cylinder's surface
are implemented using the method of images, which enforces the  
condition by ensuring that the boundary is a streamline. 
An image of the vortex line configuration is generated with 

\begin{equation}
\mathbf{s}_{\text{image}}=\left(
\frac{a^2 s_x}{s_x^2+s_y^2},\,\frac{a^2 s_y}{s_x^2+s_y^2},\,s_z\right),
\end{equation}

\noindent
where $(s_x, s_y, s_z)$ are the $(x,y,z)$
components of the Lagrangian discretization point $\mathbf{s}$ 
of the original vortex line. 
Similarly to the reconnection procedure used in the bulk of the
fluid when two vortex lines collide, vortex lines that pass within 
a distance of the order of the minimum separation from
the surface of the cylinder are reconnected algorithmically 
to their images.

\begin{figure}[ht]
\begin{center}
\includegraphics[width = 0.80\linewidth]{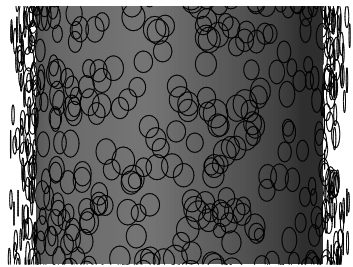}
\caption{Typical initial vortex configuration of vortex rings
in the vicinity of the cylinder.}
\label{fig1}
\end{center}
\end{figure}

The initial state consists of vortex rings placed
in the immediate vicinity of the cylinder's surface, whose orientations 
are chosen such that the initial velocity of each ring is in the outward 
radial direction, see Figure~\ref{fig1}.
The initial radii of the rings are drawn from a normal 
distribution with mean $0.007~\rm~cm$ and standard deviation 
$0.001~\rm cm$. We have also experimented with simpler
initial vortex configurations, such as randomly oriented vortex rings, 
finding that they tend to
require a longer transient to achieve a statistically identical turbulent steady-state.

\section{Absence of steady-state for uniform temperature}
\label{sec:uniform}

\begin{figure}[ht]
\begin{center}
\includegraphics[width = 0.96\linewidth]{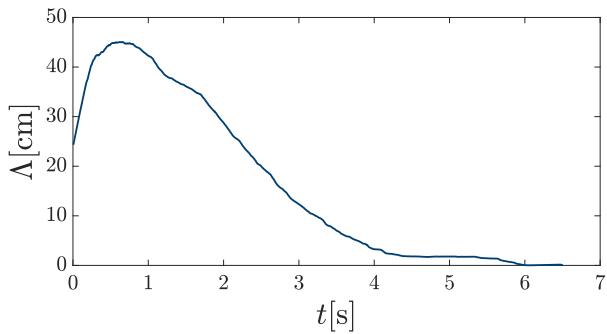}
\caption{The total vortex line length $\Lambda$ vs time in the case 
where the temperature and mutual friction coefficients are uniform 
throughout the entire flow domain. Notice that the turbulence decays.
}
\label{fig2}
\end{center}
\end{figure}

First we consider the case in which the temperature is uniform
throughout the whole flow domain $r>a$ with value $T=1.3~\rm K$.
As it can be seen from Figure~\ref{fig2}, which illustrates the time  
evolution of the total vortex line length $\Lambda$ within the whole flow 
domain, the vortex tangle does not saturate to a statistically steady state.
We find that at first $\Lambda$ increases, and then crashes to zero, 
independently of the initial condition used.
This result (the absence of a steady-state),
which we obtain both under an initial LIA evolution, as described
in the previous section, and also under the Biot-Savart law from
start to finish,  is consistent with recent 
calculations by Varga~\cite{Varga2019} who simulated spherically symmetric 
counterflow using the same vortex filament method. 
In a recent paper \cite{Sergeev-EPL}, we have explained the absence
of a steady-state solution using the 
Hall-Vinen-Bekarevich-Khalatnikov (HVBK) 
equations~\cite{Khalatnikov-book}, i.e.
a continuous model of the laminar vortex flow as well as the turbulent
flow \cite{Roche2009} of helium~II.
According to the HVBK equations, if 
the temperature is uniform and the helium's properties are
constant in the flow domain, then there exists only {\it one} steady-state
solution corresponding to a particular choice of $\vo$, that is, a single
value of heat flux. For an arbitrary heat flux, there is no steady-state
solution.

\section{Turbulent steady-state for nonuniform friction}
\label{sec:steady-state}

According to Ref.~\cite{Sergeev-EPL},
the HVBK equations allow steady-state 
solutions only in the case where the temperature in the bulk of helium 
is no longer assumed constant, but is found as a function of radial coordinate 
from the equations; this means that the normal fluid and
superfluid densities as well as all other thermodynamic variables
and the mutual friction coefficients depend on the radial coordinate.
Unfortunately it would be computationally and conceptually
difficult to include this spatial
variability in the vortex filament method (for example,
Eq.~(\ref{eq:BS}) assumes that the superfluid is incompressible).
At the same time, in order to make progress in this problem,
it would be useful to go beyond the HVBK equations and their
limitations when applied to turbulence (if the vortex lines are 
randomly oriented, the superfluid vorticity
is not related to the vortex line density 
\cite{BaggaleyLaurie2012} in the simple way assumed by the HVBK equations).

\begin{figure}[ht]
\centering
\begin{tabular}{cc}
\includegraphics[width = 0.96\linewidth]{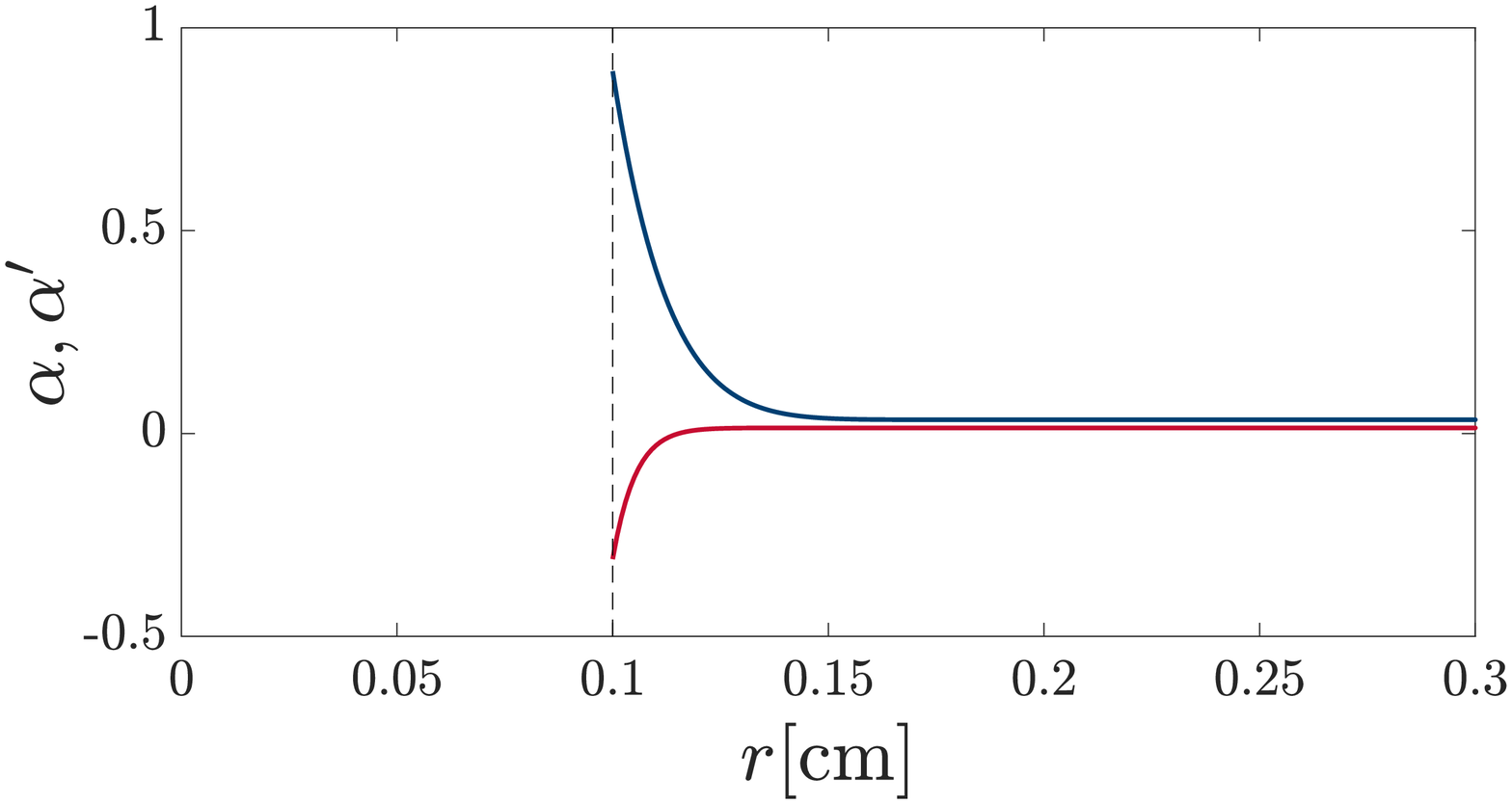} \\
\includegraphics[width = 0.96\linewidth]{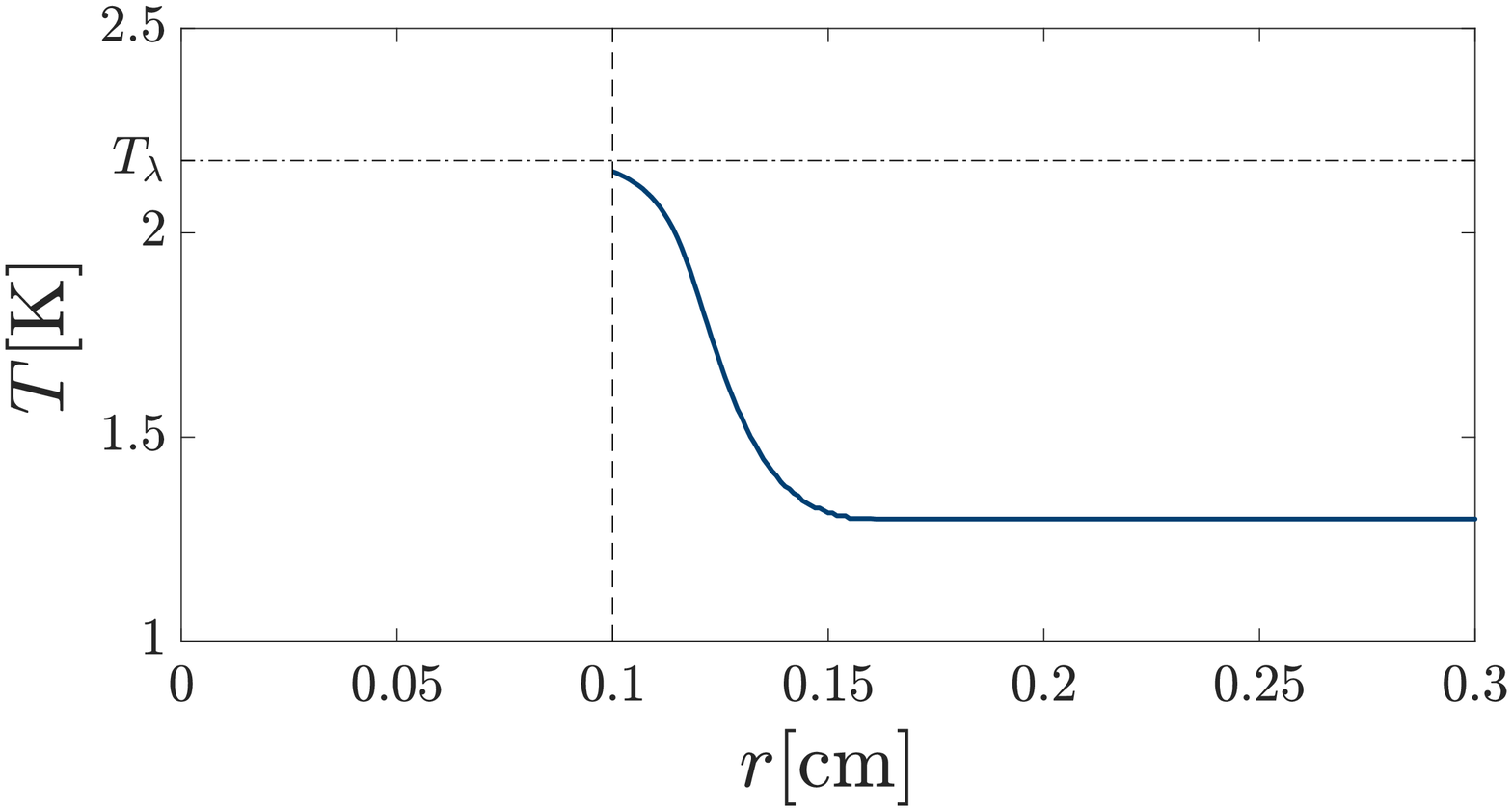}
\end{tabular}
\caption{Top panel: mutual friction coefficients $\alpha$ (top, blue) 
and $\alpha'$ (bottom,red) as functions of the radial coordinate. 
Bottom panel: radial profile of temperature corresponding to the 
distributions of mutual friction coefficients shown in the top panel.}
\label{fig3}
\end{figure}

Based on these motivations and our previous findings 
\cite{Sergeev-EPL} here we develop a minimal numerical 
model of turbulent radial counterflow which captures the
essential physics of the problem and lets us use the vortex
filament model (the best model available for turbulent
helium~II at nonzero temperatures): we assume that the normal 
fluid and superfluid densities are constant throughout the 
whole flow domain, 
but the mutual friction coefficients vary with the radial coordinate, 
that is $\alpha=\alpha(r)$ and $\alpha'=\alpha'(r)$. 
Furthermore, we assume that the behavior of $\alpha$ and $\alpha'$ 
mimics the  radial profiles of these coefficients 
that follow from a typical radial distribution of temperature 
described in Ref.~\cite{Sergeev-EPL}. 
The radial distributions of $\alpha$ and $\alpha'$ used in our numerical 
simulations are shown in the top panel of 
Figure~\ref{fig3}.

These distributions conform to $\alpha(r)$ and 
$\alpha'(r)$ corresponding to the case where the surface 
temperature is $\To=2.15$~K, and the bulk temperature of helium 
is $1.3 \rm~K$. Note that the mutual friction coefficients undergo 
a rapid change only within a relatively narrow region (about half a 
radius $a$ from the cylinder's surface) before saturating to constant 
values in the bulk of helium. The radial profile of temperature 
corresponding to the distributions of $\alpha$ and $\alpha'$ shown in 
the top panel of Figure~\ref{fig3} is illustrated in the bottom 
panel of this figure.

\begin{figure}[ht]
\begin{center}
\includegraphics[width = 0.96\linewidth]{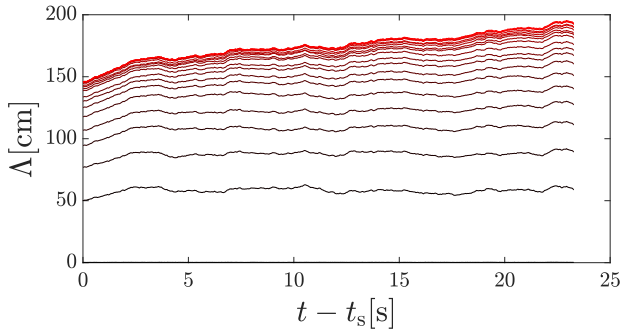}
\caption{Evolution of the total line length, $\Lambda_n$ vs time 
(starting from the time $t_s$ defined in the text) within 
cylindrical shells, of outer radii increasing with $n$, 
adjacent to the cylinder's surface, see text for details. 
The curves, from bottom to top, correspond to $n=1,\,2,\,3,\,\dots$.}
\label{fig4}
\end{center}
\end{figure}

Under the assumptions of our minimal model,
the saturation of the tangle to a statistically steady-state is 
illustrated by Figure~\ref{fig4} which shows the total vortex line 
length, $\Lambda_n$ 
($n=1,\,2,\,3,\,\dots$) vs time (starting from the time $t_s$)
within cylindrical shells whose inner 
radius is that of the cylinder, $r=a=0.1 \rm~cm$, and the outer radius 
of the $n$th shell is $(n+1)a$. The curves, from bottom to top, 
correspond to $n=1,\,2,\,3,\,\dots$. 
The top red line shows the total 
line length within the whole simulation domain. 
Figure~\ref{fig4} clearly shows the trend to saturation 
(within each shell) of the vortex line density to a statistically steady state. 
It can be seen that saturation is achieved
very quickly within the first shell, and 
in about $5$ - $10~\rm s$ from $t_s$
within the fifth shell.
Times of saturation become longer within shells of bigger outer radii, 
as the outer regions may contains some large
vortex loops which may slowly move away.
It is surprising that the saturation of the vortex tangle 
throughout the whole flow domain is the consequence of the radial 
dependence of the mutual friction coefficients within a rather 
small region, just about half a cylinder's radius from the surface.
Figure 4 also illustrates the criterion for switching from LIA to
Biot-Savart: $t_s$ is the time when $\Lambda \approx 120~\rm cm$ 
in the fifth shell, corresponding to a turnover time $\tau \approx 1.8~\rm s$
which is much shorter than the computed Biot-Savart evolution.

\begin{figure}[ht]
\begin{center}
\includegraphics[width = 0.94\linewidth]{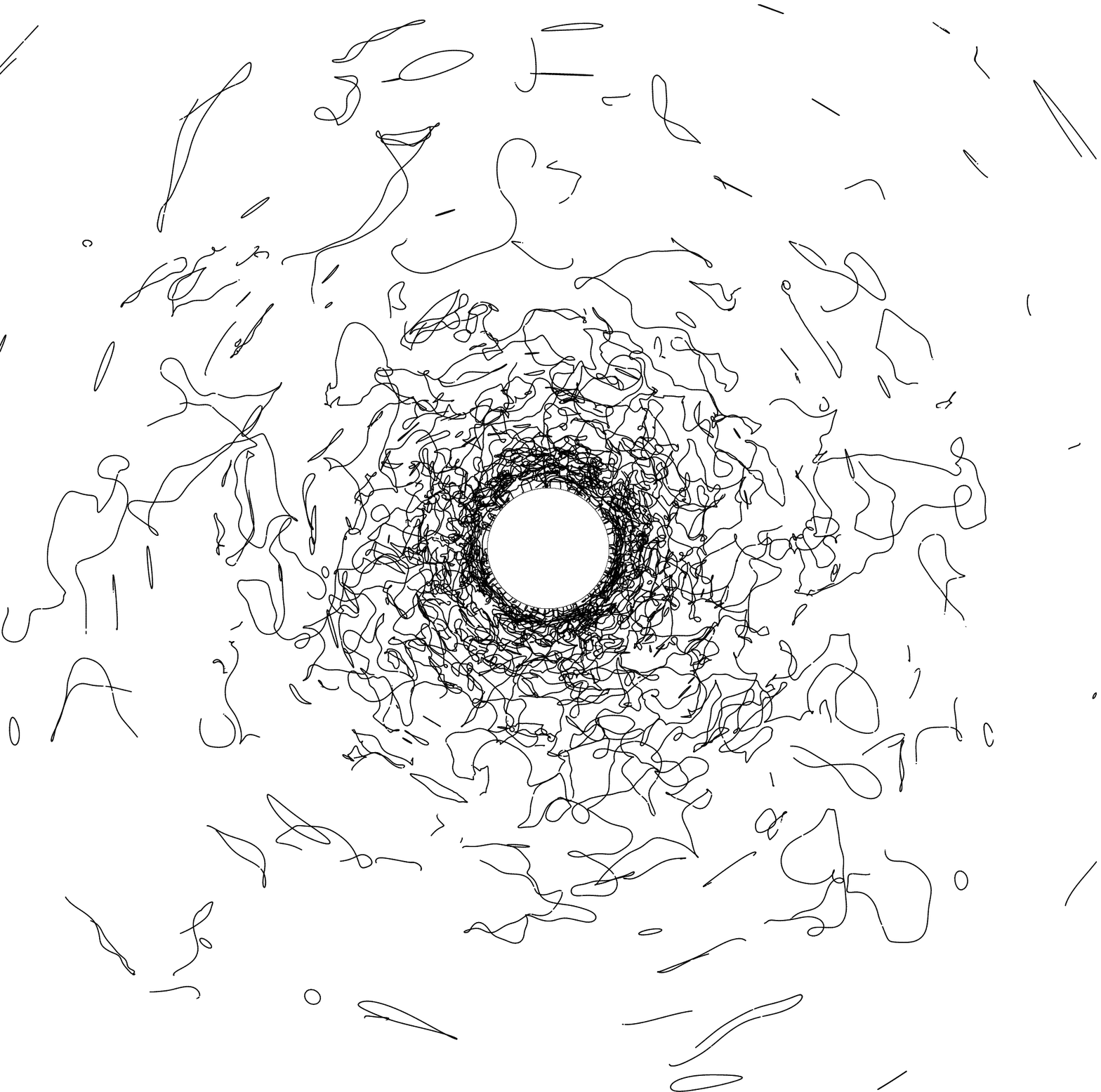}
\includegraphics[width = 0.96\linewidth]{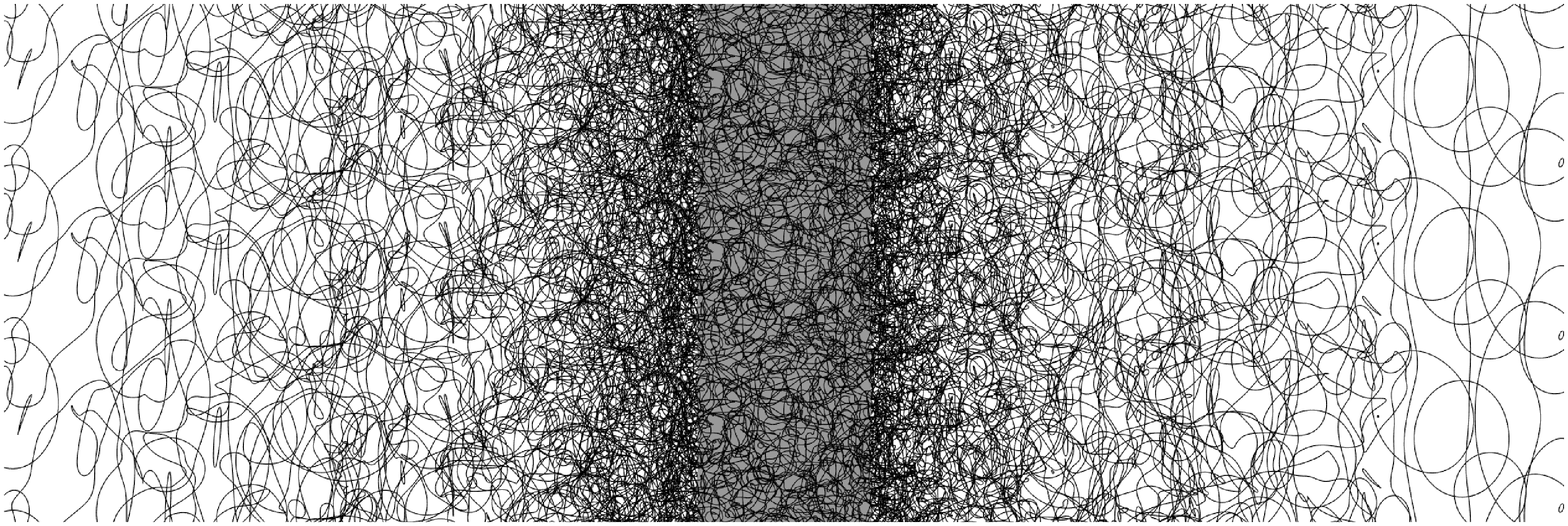}
\caption{Configuration of quantized vortex lines 
in the statistically steady-state regime. 
Top: top view, bottom: side view.}
\label{fig5}
\end{center}
\end{figure}

Figure~\ref{fig5} illustrates the numerically simulated,
statistically steady-state vortex tangle. 
It is interesting to notice that whereas in the inner regions
the tangle is dense and apparently random, in the outer region 
large irregular vortex loops are visible
which are oriented in the plane
perpendicular to the radial direction of the heat flux. This effect
can be understood in terms of the the dynamics of simpler
circular vortex rings. Starting from the dense region, a vortex ring
which travels in the outward radial direction gains energy from the
normal fluid which flows in the same direction (radially out), slows down and
becomes larger (thus gaining energy), 
until it collides with similar large loops, forming
the large stationary vortex structures  
visible at the edge of Figure~\ref{fig5}(top) near $r \approx 8 a$.

\begin{figure}[ht]
\begin{center}
\includegraphics[width = 0.96\linewidth]{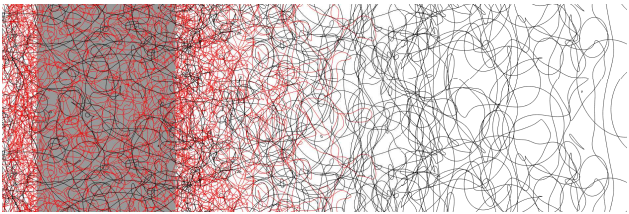} 
\caption{
Enlarged view of the vortex configuration near the cylinder. 
Vortex handles
(which are connected to the cylinder) are plotted in red,
vortex loops (which are disconnected) are plotted in black.}
\label{fig6}
\end{center}
\end{figure}

\begin{figure}[ht]
\begin{center}
\includegraphics[width = 0.96\linewidth]{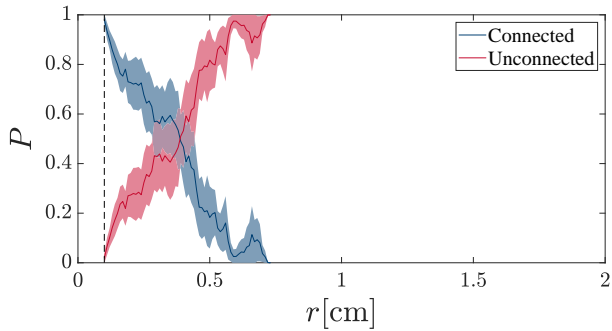}
\caption{
Relative proportion, $P$, of vortex loops (disconnected from the
cylinder, pink)
and vortex handles (connected to the cylinder, blue) 
as a function of radial distance, $r$.
The coloured bands represent one standard deviation after time-averaging
in the statistically stead-state regime.}
\label{fig7}
\end{center}
\end{figure}

To gain geometrical insight into the turbulence,
at each instant of time we can
divide the vortex configuration of vortex lines in two groups
depending whether they are closed or attached to the boundary:
"vortex handles" (which are connected to the cylinder), and
"vortex loops" (distorted vortex rings)
which are disconnected from the cylinder.
Figure~\ref{fig6} displays the former in red and the latter in black.
Figure~\ref{fig7} shows that the relative proportion of vortex handles
is larger near the cylinder, whereas vortex loops
are predominant far from the cylinder.

\begin{figure}[ht]
\centering
\begin{tabular}{cc}
\includegraphics[width = 0.96\linewidth]{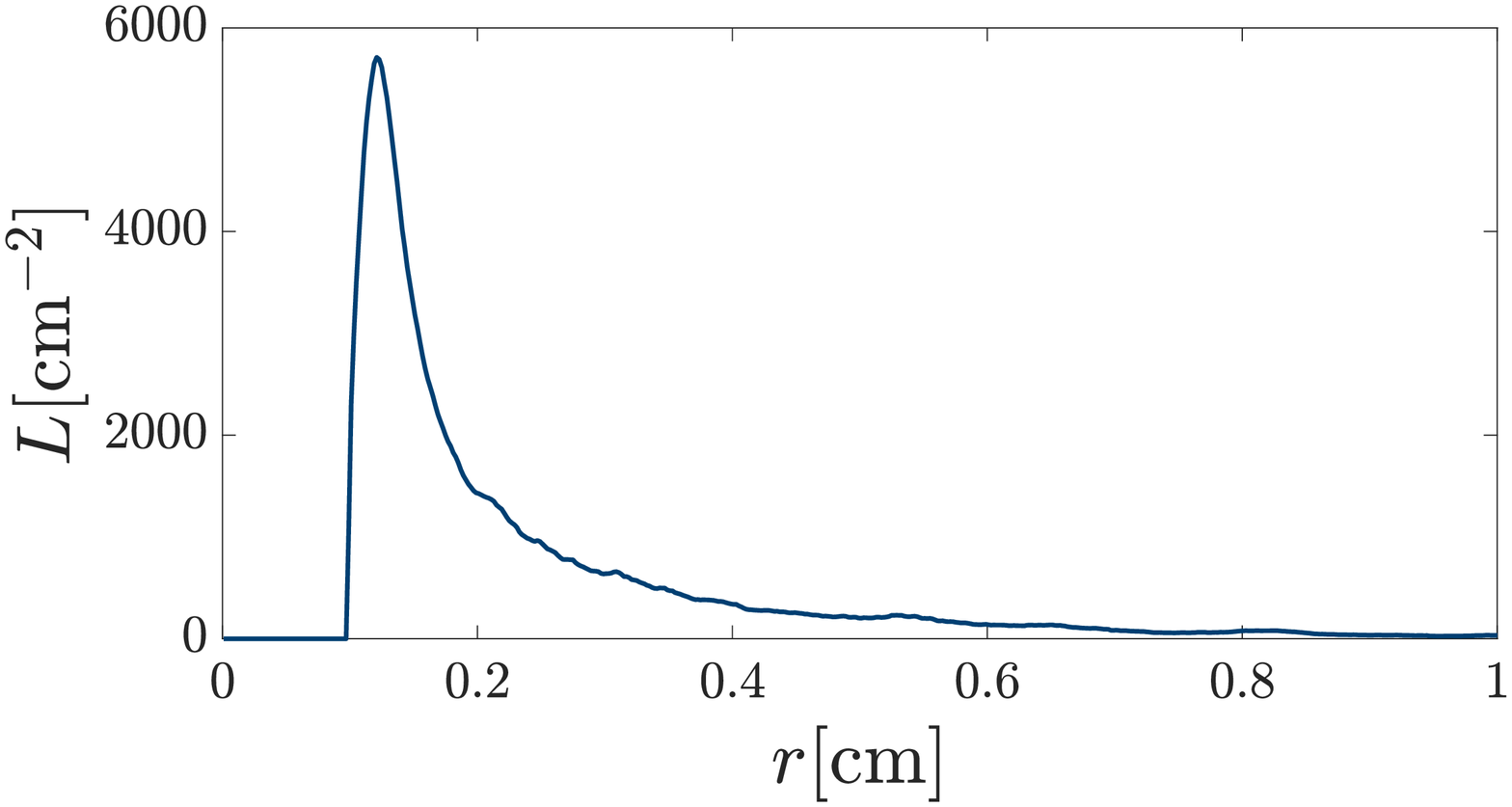} \\
\includegraphics[width = 0.96\linewidth]{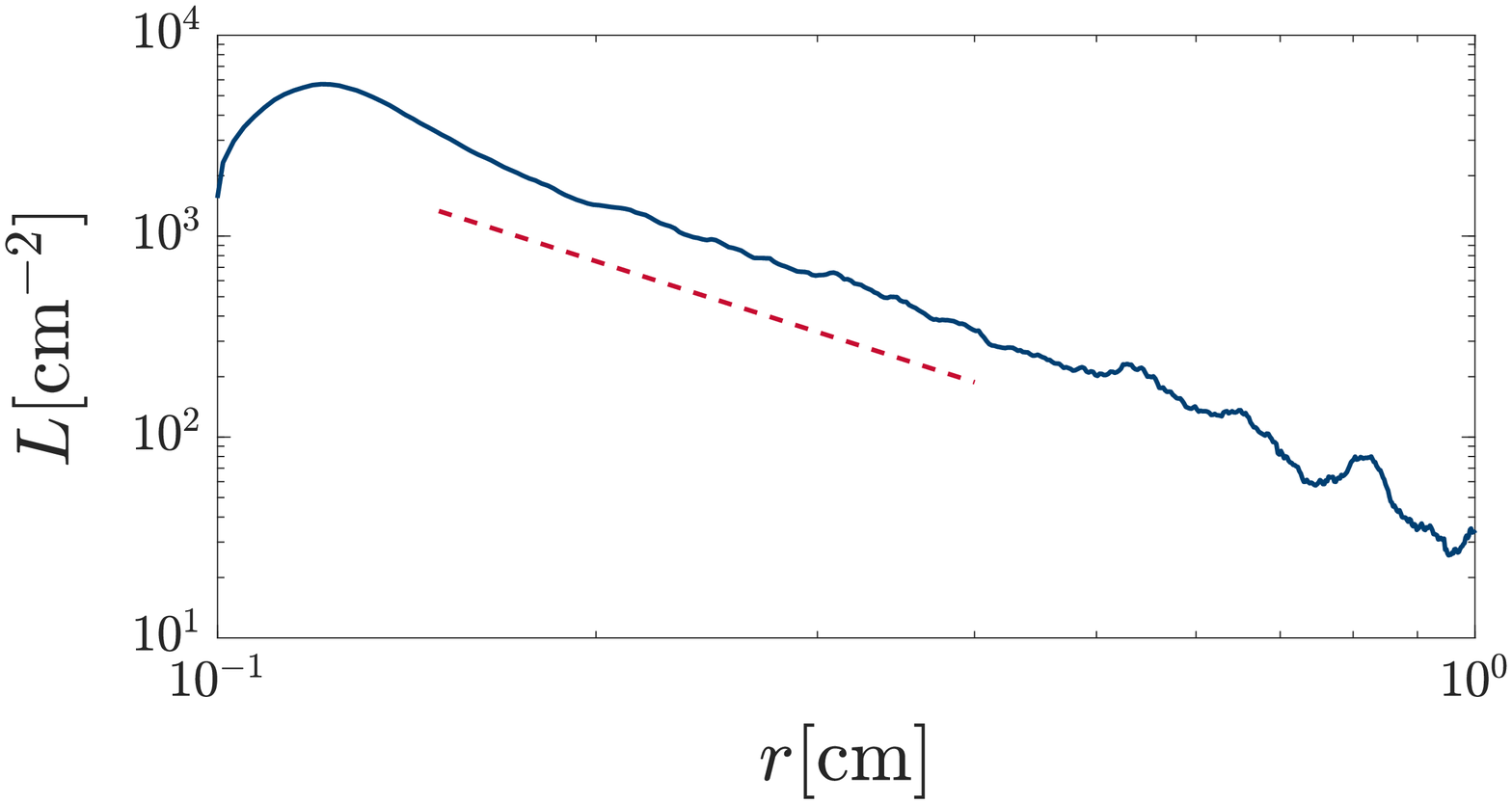}
\end{tabular}
\caption{Top: radial distribution of the vortex line density $L(r)$
in the flow domain $r>a=0.1~\rm cm$ where $a$ is the radius of the 
cylinder. Bottom: $L(r)$ versus $r$ plotted on the log-log scale; 
the dashed (red) straight line shows the expected $r^{-2}$ scaling.}
\label{fig8}
\end{figure}

It is interesting to analyse the properties of radial counterflow 
turbulence and compare them
quantitatively to traditional counterflow turbulence in a channel.
From our numerical solution we calculate the local average vortex 
line density in the statistically steady state regime as follows: 
we divided the flow domain for $0.1~{\rm cm}\leqslant r\leqslant1~{\rm cm}$ 
into thin cylindrical shells of thickness $0.005 \rm~cm$. 
Within each shell the vortex line density is then ensemble-averaged 
over a number of realizations of the initial vortex configuration. 
Figure~\ref{fig8} shows the result with the top panel plotted
in the linear-linear scale and the bottom panel in the log-log scale.
The large radial inhomogeneity of radial counterflow
turbulence is apparent.

If the temperature is constant and the flow is steady,
in the framework of the HVBK equations, 
the mass conservation equations reduce to $\bnabla\cdot\vvs=0$ 
and $\bnabla\cdot\vvn=0$. In the case of cylindrically symmetric 
radial counterflow these equations yield the radial distributions of 
the normal and superfluid velocities which scale with the radial 
coordinate as $r^{-1}$. Then, from the well-known Gorter-Mellink 
relation

\begin{equation}
L=\gamma^2\vns^2,
\label{eq:GM}
\end{equation}

\noindent
where $\vns=\vert\vvn-\vvs\vert$ is the counterflow velocity 
and $\gamma=\gamma(T)$ is a temperature-dependent constant 
(see e.g. Refs.~\cite{Adachi,Kondaurova}), 
we expect that the vortex line density scales as $L \sim r^{-2}$.
As seen from the bottom panel of Figure~\ref{fig8}, at distances larger 
than about half a cylinder's radius from the surface,
the vortex line density behaves reasonably close to $\sim r^{-2}$ indeed,
although some deviation from this scaling is apparent. 
This deviation is most likely due to the well-known observation 
that Eq.~(\ref{eq:GM}) should also include the 
so-called intercept velocity, $v_i$, and hence be written 
in a slightly different form~\cite{Tough}:

\begin{equation}
L=\gamma^2(\vns-v_i)^2.
\label{eq:GMi}
\end{equation}

\begin{figure}[ht]
\begin{center}
\includegraphics[width = 0.96\linewidth]{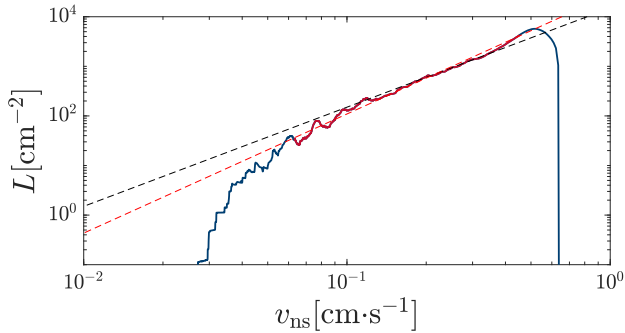}
\caption{Vortex line density as a function of the counterflow velocity. 
The black dashed line shows the scaling $L\propto\vns^2$, in good
agreement with our data. 
The red dashed line shows the best fit\cite{note2}, which is
$L \propto \vns^{2.4}$.}
\label{fig9}
\end{center}
\end{figure}

\noindent
The intercept velocity is typical of turbulence 
at low counterflow velocities, 
corresponding to rather dilute vortex tangles \cite{Tough}.
Calculated from our numerical solution, the local average vortex line density 
as a function of the radially dependent counterflow velocity 
$\vns=\vn-\vs$ is shown in Figure~\ref{fig9}. 
Within the interval of counterflow velocities
roughly corresponding to the flow region where the line density is 
fully saturated, 
the behavior of the vortex line density is close \cite{note2}
to the expected scaling $L\propto\vns^2$, although some deviation from 
this scaling is apparent at low velocities. 
Note that our numerical results allow us to estimate the intercept 
velocity as $v_i\approx 0.002~\rm cm/s$, a value that is one order 
of magnitude smaller than that reported in Ref.~\cite{Kondaurova} 
(see also references therein) for counterflow in straight channels.

Finally, we measure the anisotropy of the turbulence
as a function of radius using the anisotropy parameters of Schwarz 
\cite{Schwarz1988}:

\begin{equation}
I_{\parallel}=\frac{1}{\Lambda}\int_{\mathcal{L}}\left[1-(\mathbf{s}^{\prime}\cdot\hat{\mathbf{r}}_{\parallel})^2\right]d\xi,
\end{equation}

\begin{equation}
I_{\perp}=\frac{1}{\Lambda}\int_{\mathcal{L}}\left[1-(\mathbf{s}^{\prime}\cdot\hat{\mathbf{r}}_{\perp})^2\right]d\xi,
\end{equation}

\noindent
where $\hat{\mathbf{r}}_{\parallel}$ and $\hat{\mathbf{r}}_{\perp}$ 
are unit vectors, respectively parallel and 
perpendicular to the direction of the counterflow velocity.
Schwarz's anisotropy parameters satisfy $I_{\parallel} /2 + I_{\perp}=1$.
If the vortex lines are aligned in the plane perpendicular to the counterflow 
direction, then
$I_{\parallel}=1$ and $I_{\perp}=1/2$, while if the vortex lines 
are isotropic $I_{\parallel}=I_{\perp}=2/3$. Simulations of counterflow
turbulence in a periodic box at $T=1.3~\rm K$
report \cite{Adachi} $I_{\parallel} \approx 0.74$ in agreement
with counterflow channel experiments ($I_{\parallel} \approx 0.77$).

\begin{figure}[ht]
\begin{center}
\includegraphics[width = 0.98\linewidth]{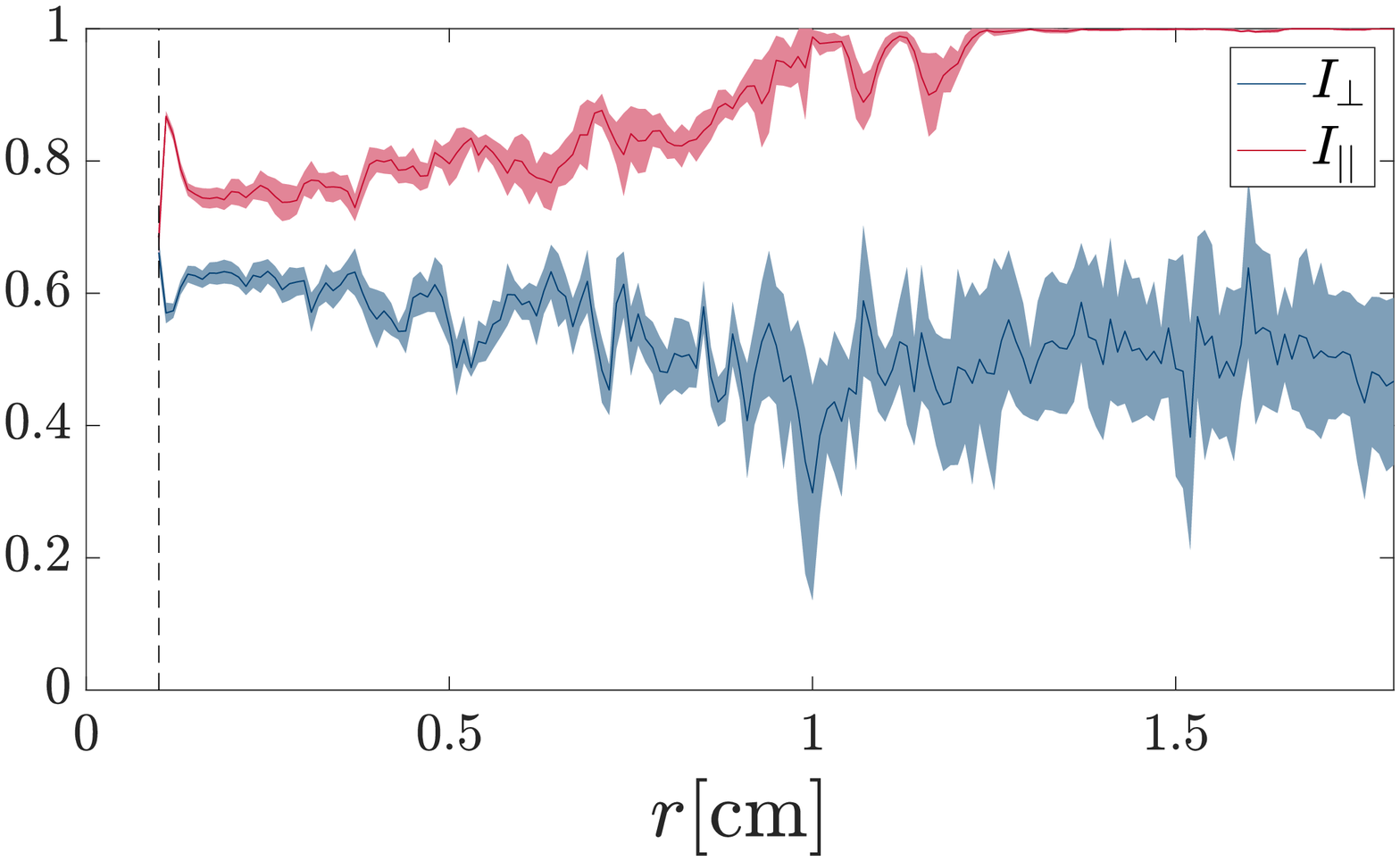}
\includegraphics[width = 0.98\linewidth]{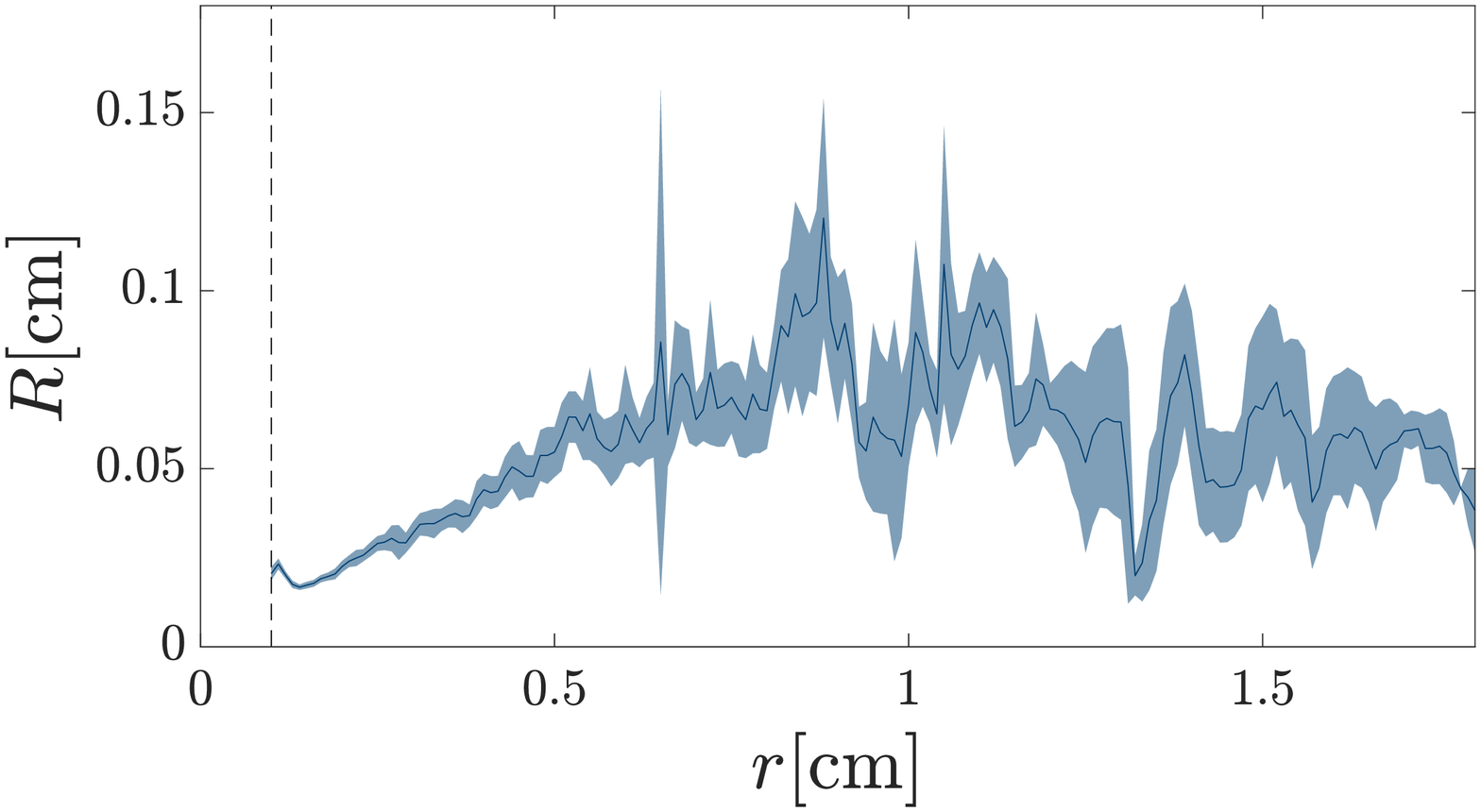}
\caption{
Top panel: anisotropy parameters of Schwarz \cite{Schwarz1988},
$I_{\parallel}$ (bottom, blue) and $I_{\perp}$ (top, pink),
as a function of radial distance, $r$; the coloured bands represent
one standard deviation. Bottom panel: average radius of curvature 
of the vortex lines, $R$, as a function of $r$; again the coloured
band represents one standard deviation}.
\label{fig10}
\end{center}
\end{figure}

In our case $\hat{\mathbf{r}}_{\parallel}$ is the outward radial unit
vector.
$I_{\parallel}$ and $I_{\perp}$ are estimated locally by calculating them 
over radial shells of width $0.01$ cm, with $\Lambda$ then 
being the total vortex line length within the shell and $\mathcal{L}$ 
being the vortex lines within the shell. 
Figure~\ref{fig10}(top)
shows that the turbulence is not isotropic anywhere. In the inner region
very near the cylinder ($r \geq a$) and in the
outer region ($r \geq 7a$), most vortex length lies in the plane 
perpendicular to the radial direction of the counterflow; in the
intermediate region the anisotropy of radial counterflow is comparable
to what is observed in standard channels.
Figure~\ref{fig10}(bottom)
shows the average radius of curvature, $R$, of the vortex loops, defined
locally as $R=1/\vert \sss''\vert$. This geometrical property determines
the dynamics (large vortex loops move slower than small loops). The graphs
shows clearly that away from the cylinder there is a region dominated
by large slow-moving vortex loops,
consistently with Figure~\ref{fig5}.

\section{Conclusions}
\label{sec:conclusions}

In our previous work~\cite{Sergeev-EPL} we have shown that,
in the framework of the HVBK model applied to turbulence, 
the governing equations do not have a steady radial counterflow 
solution if the temperature of helium is assumed uniform throughout 
the whole flow domain.  We have also shown that 
the vortex line density saturates to a steady state value only in the case 
where the radial distributions of temperature, normal fluid
and superfluid densities, thermodynamic properties, and 
mutual friction coefficients are accounted for. 

Using these ~\cite{Sergeev-EPL} findings, here we have developed
a minimal model of turbulent radial counterflow
which identifies the essential physics of the problem from
the point of view of the vortex dynamics: the
radial dependence of the friction. Our model
assumes that the mutual friction coefficients are
radially dependent and leaves the normal and superfluid densities 
and all thermodynamic properties constant throughout the flow domain.
This radial dependence of the mutual friction coefficients
(which mimics the changes calculated from the HVBK 
equations~\cite{Sergeev-EPL}) takes place only within a relatively 
narrow region, about a half of the cylinder's radius adjacent to 
the cylinder's surface; outside this region the mutual friction coefficients
retain constant values corresponding to the temperature in the bulk of helium. 
With this minimal model, the numerical simulations of turbulent  
counterflow based on the Biot-Savart law which we 
present here show that a dense layer of turbulent vortex lines forms
near the heated cylinder and saturates to a statistical steady-state.

We have analysed the inhomogeneous vortex line density $L$ of this
turbulent steady state and determined its anisotropy and its scalings
with the radial coordinate, $r$, and with the counterflow 
velocity, $\vns$. We have found scalings which are reasonably close to  
$L \propto r^{-2}$ and $L \propto \vns^2$ expected from
homogeneous counterflow turbulence in standard channels. 

At present, a numerical model based on the Biot-Savart law
that goes beyond the current minimal model and properly
incorporates spatial variations of temperature together with 
temperature-dependent densities, thermodynamic properties, and 
mutual friction coefficients seems too ambitious, considering 
limitations of computing power. However, the identification of
the spatial dependence of the mutual friction
as the main physical mechanism responsible for the 
saturation of the vortex tangle to a statistically steady state,
which is the main result of this paper,
should help the study of thermal counterflow in the case of 
other non-trivial flow geometries. Moreover, the validity of
Vinen's scaling $L \propto \vns^2$ in strongly inhomogeneous
turbulence which we have directly verified in this work is going 
to help develop further the theory of inhomogeneous superfluid turbulence.

\bigskip

We acknowledge the support of the EPSRC (Engineering and Physical Sciences 
Research Council) grant EP/R005192/1.


\begin{thebibliography}{99}
\bibitem{note0}
The results which we present here do not apply directly
to current anemometry experiments, as in these experiments
the temperature may exceed
$T_{\lambda}$ and in the lambda region the vortex line density would
become to large for our simulations.
\bibitem{note1}
More precisely, to define $t_s$ we monitor the vortex length within
the fifth shell around the cylinder, as defined in the caption
of Figure~\ref{fig4}.
\bibitem{note2}
A slightly better fit (shown by the red dashed line in Figure~\ref{fig9})
is provided by $L\propto(\vns-v_i)^{2.4}$ with $v_i\approx0.006$~cm/s 
corresponding to the region $0.06< \vns <0.8~\rm cm/s$.
\end{thebibliography}
\end{document}